\documentclass[final,1p,times]{elsarticle} % do not change
\usepackage{graphicx} % do not change
\usepackage{amssymb} % do not change
\usepackage{fixmath} % bold math headlines          
\usepackage{bm} % bold math          
\usepackage{amsthm} % do not change
\usepackage{lineno} % do not change

\newcommand{ \be }{\begin{eqnarray}}
\newcommand{ \ee }{\end{eqnarray}}
\newcommand{ \la }{\langle}
\newcommand{ \ra }{\rangle}

\newcommand{ \mean }[1]{\left\langle #1 \right\rangle}   

\def\snn{$\sqrt{s_{NN}}$}

\def\P{$\cal P$}

\newcommand{ \psirp }{\Psi_{RP}}
\newcommand{ \phia }{\phi_{\alpha}}
\newcommand{ \phib }{\phi_{\beta}}
\newcommand{ \corr }{\mean{\cos(\phia+\phib-2\psirp)}}   

\usepackage{color}

\journal{Nuclear Physics A} % do not change
\begin{document} % do not change

\begin{frontmatter} % do not change

\title{Experimental study of local strong parity violation in
relativistic nuclear collisions}

\author{Sergei A. Voloshin and the STAR Collaboration}

\address{Department of Physics and Astronomy\\
Wayne State University, Michigan 48201 USA }

%\ead{voloshin@wayne.edu}

\begin{abstract}
Parity-odd domains, corresponding to non-trivial topological solutions
of  the QCD  vacuum, might  be  created in relativistic heavy  ions
collisions.  These domains are  predicted to lead to charge separation
along  the  system orbital momentum  of the system
created in non-central collisions.    
Three-particle  mixed  harmonics azimuthal correlator is a
\P~even observable  but
  directly  sensitive  to  the  charge  separation  effect.
Using this observable to analyze Au+Au and Cu+Cu collisions at
$\sqrt{s_{NN}}=200$ and 62~GeV, STAR detects
a signal  consistent  with several of the
theoretical expectations.    
Possible contributions  from effects not 
related to  parity  violation are studied with existing event
generators, which fail to describe the data.
Future directions in studying the effect are discussed.  
\end{abstract}

\end{frontmatter} % do not change

%\maketitle
%% QM09: we keep linenumbers at least for initial version
% \linenumbers % do not change

%%%%%%%%%%%%%%%%%%%%%%%%%%%%%%%%%%%%%%%%%%%%%%%%%%%%%%%%%%%%%%
{\bf \textsl{1. Introduction.}}
Quantum Chromodynamics (QCD) is the theory of strong interactions. 
Perturbative QCD is firmly established and thoroughly tested experimentally. 
In the non-perturbative sector, QCD links chiral symmetry breaking and the 
origin of hadron masses to 
the existence of topologically non-trivial classical solutions 
describing the transitions between the vacuum states with 
different Chern-Simons numbers.  
Quark interactions with topologically non-trivial classical gluonic fields
 change the quark helicity and are $\cal P$~and $\cal CP$~odd. 
It was suggested in~\cite{Kharzeev:1998kz} 
that metastable 
$\cal P$~and $\cal CP$~odd domains,                 
characterized by non-zero topological charge, might be created in 
ultra-relativistic heavy ion collisions. 
The possibility for an experimental detection of this  
{\em  local strong  parity  violation}
was discussed
in~\cite{Kharzeev:1998kz,Voloshin:2000xf,Finch:2001hs}.
More recently, it was noticed~\cite{Kharzeev:2004ey,Kharzeev:2007tn} 
that in non-central collisions such domains 
can demonstrate themselves via the asymmetry in the emission 
of charged particle
w.r.t. the system's angular momentum.
Such charge separation is a consequence of the difference in the number
of particles with positive and negative helicities positioned
in the strong magnetic field ($\sim 10^{15}$~T) 
of  a non-central nuclear collision~\cite{Kharzeev:2004ey,Kharzeev:2007jp},
the so-called {\em chiral magnetic effect}.
The same phenomenon can also be described in terms of the induced electric
field that is parallel to the static external magnetic field, 
which occurs in the presence of topologically non-trivial vacuum 
solutions~\cite{Fukushima:2008xe}. 

Since the direction of the separation may vary event by 
event in accord with the changing sign of the topological charge of the domain,
the observation of the effect is possible only by correlation techniques.
Such an observable, $\cal P$-even, but directly sensitive to the charge
separation effect, has been proposed in~\cite{Voloshin:2004vk}
and is based on 3-particle mixed harmonics azimuthal correlations.

Phenomenologically, the charge separation can be  described 
by adding a \P-odd sine term to the Fourier decomposition of 
the particle azimuthal distribution 
with respect to the reaction plane angle, $\Psi_{RP}$,
which is often used in the description of
anisotropic flow~\cite{Voloshin:2004vk}: 
\be
 \frac{dN_\pm}{d\phi} &\propto& 1 + 2 v_1 \cos(\Delta \phi)+
2 v_2 \cos(2\Delta\phi)+...
%\nonumber \\
+ 2 a_{\pm} \sin(\Delta \phi) +... \, ,
\label{eq:expansion}
\ee
where $\Delta \phi =(\phi-\psirp)$ is the particle azimuth 
relative to the reaction plane,
$v_1$ and $v_2$ account for directed and elliptic flow. 
Parameters $a_- = -a_+$ describe the \P-violating effect.
The sign of $a_\pm$ 
varies event to event following the fluctuations in 
the domain's topological charge, and on average is zero, $\mean{a_\pm}=0$. 
The observation of the effect is possible via correlations,
e.g. $\mean{a_\alpha a_\beta}$,
where $\alpha$ and $\beta$ denote the particle type. 
To measure $\mean{a_\alpha a_\beta}$, 
it was proposed~\cite{Voloshin:2004vk} to use the correlator:
\be
\hspace*{-2cm}
& \mean{ \cos(\phia +\phib -2\psirp) } = 
\label{eq:obs1}
%\\  =
\mean{\cos\Delta \phia\, \cos\Delta \phib} 
-\mean{\sin\Delta \phia\,\sin\Delta \phib}
\label{eq:cossin}
\\ 
& =
[\mean{v_{1,\alpha}v_{1,\beta}} + B_{in}] - [\mean{a_\alpha a_\beta}
+ B_{out}] \approx - \mean{a_\alpha a_\beta} + [ B_{in} - B_{out}].
\label{eq:v-a}
\ee
This correlator represents the
difference between correlations projected onto an axis in
the reaction plane and the correlations projected onto an axis
perpendicular to the reaction plane. 
The key advantage of using such a difference is that it
removes all the correlations among
particles $\alpha$ and $\beta$ that are not related to the reaction plane 
orientation~\cite{Borghini:2002vp,Adams:2003zg}.
Only the parts of such correlations that depend on azimuthal orientation 
with respect to the reaction plane remain as backgrounds, denoted as
$[B_{out}-B_{in}]$.  
Note that the contribution given by the term $\mean{v_{1,\alpha}v_{1,\beta}}$
can be neglected because directed flow averages to zero in
a rapidity region symmetric with respect to mid-rapidity, as used in this
analysis.

According to Refs.~\cite{Kharzeev:2004ey,Kharzeev:2007tn,Kharzeev:2007jp}
one expects the following features of the correlator
$\mean{a_\alpha a_\beta}$:
\begin{itemize}
\item 
{\em Magnitude}: Estimates~\cite{Kharzeev:2004ey}
 predicted a signal 
 $|a|\sim Q/N_{\pi^+}$, where $Q=0,\pm 1,\pm 2, ...$ 
is the topological charge and
 $N_{\pi^+}$ is the positive pion multiplicity in 
one unit of rapidity 
-- the typical scale of such correlations.
More accurate estimates~\cite{Kharzeev:2007jp} 
including the strength of the magnetic
 field and topological domains production rates, were found to be close to
 the same number, of the order of $10^{-2}$ for mid-central collisions
at top RHIC energies,
which corresponds to $10^{-4}$ for the correlator $\mean{a_\alpha a_\beta}$.
\item
{\em Charge combinations}: 
In the absence of medium effects, one expects 
$\mean{a_+a_+}=\mean{a_-a_-}=-\mean{a_+a_-} >0$. 
In a dense medium 
suppression of the back-to-back correlations may occur~\cite{Kharzeev:2007jp}:
$\mean{a_+a_+}=\mean{a_-a_-} \gg -\mean{a_+a_-}$. 
The effect of strong radial flow can further modify this relation such
that the opposite charge correlations can become even positive.
\item
{\em Centrality dependence:} 
The correlator
is expected to follow a $1/N$ dependence (typical for any
kind of correlations due to clusters; $N$ is the multiplicity) 
multiplied by a factor
accounting for the variation of the magnetic field.     
Thus at large centralities the effect should decrease with centrality 
somewhat faster than $1/N$.
\item
{\em Rapidity dependence:} The correlated particles come from a 
domain of the order of 1~fm, and it is expected that the correlations
would have a typical hadronic width in $\Delta
\eta=|\eta_\alpha-\eta_\beta |$ of the order unity.
\item
{\em Transverse momentum dependence:}
Local parity violation 
is non-perturbative in nature and 
one would expect the main contribution to the signal at
$p_t \lesssim 1$~GeV/c,
but the actual limits might be affected by the radial flow.
\item
{\em  {Beam} species dependence:} 
The effect should be proportional to the square $Z^2$ of the
nuclear charge,  but the
atomic number $A$ dependence is not well understood.
The suppression of the back-to-back correlations should 
be smaller in collisions of lighter nuclei. 
\item
{\em Collision energy dependence:}
The effect might be stronger at lower energies, as the time integral
of the magnetic field is larger.
At the same time,
the charge separation effect is expected to depend strongly on
deconfinement
and chiral symmetry restoration~\cite{Kharzeev:2007jp}, 
and the signal might be greatly suppressed or completely absent
at an energy below that at which a quark-gluon plasma can be formed.
\end{itemize}

%=================================================================
{\bf \textsl{2. Data.}}
The results  are based on 14.7M Au+Au
and 13.9M Cu+Cu events at
\snn=200~GeV, and 2.4M Au+Au and 6.3M Cu+Cu events at \snn=62~GeV 
recorded with the STAR detector~\cite{Ackermann:2002ad}
 at RHIC.
Charged particle tracks were reconstructed in a 
Time Projection Chamber (TPC),  $|\eta| < 1.0$, operated
in a solenoidal magnetic field of 0.5~T.
The main TPC is supplemented 
by two Forward TPCs, which 
 cover pseudorapidity intervals  $2.7 < | \eta | < 3.9$.  
A minimum bias trigger was used during data-taking.  
The centrality of the collision is determined according to the 
recorded multiplicity of charged particles in $|\eta| < 0.5$.
The correlations are reported for charged particle in the region
 $|\eta| < 1.0$ with $p_t > 0.15$~GeV/c. 
The results integrated over transverse momentum have
 a cut of $p_t < 2$~GeV/c.

%============================================================
{\bf \textsl{3. Method.}}
\begin{figure}[h]
\begin{minipage}[t]{0.55\textwidth}
  \includegraphics[width=1.02\textwidth]{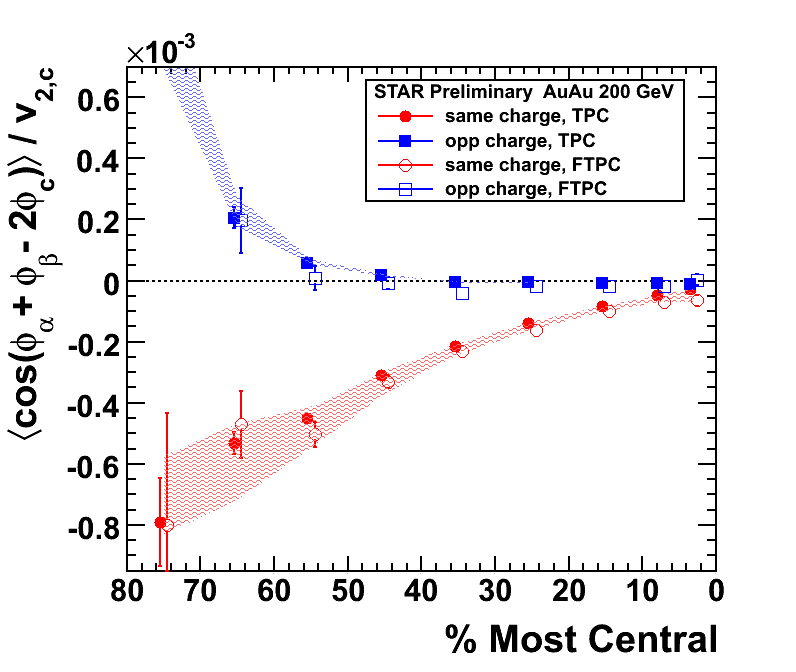}
\end{minipage}
\hspace{0.0\textwidth}
\begin{minipage}[b]{0.45\textwidth}
  \caption{
A comparison of the correlations obtained by
    selecting the  third particle  from the main  TPC or
    from  the Forward  TPCs 
    after the  results scaled by the flow of the  third  particle.
    The shaded areas represent the uncertainty from $v_{2,c}$
      scaling (see text for details).
    The error bars are statistical.
\vspace*{3mm}
}
\label{fig:au200F}
\end{minipage}
\end{figure}
In the three-particle correlation technique, used in this analysis, 
the role of the event plane
is played by the third particle that enters the  correlator 
with double the azimuth~\cite{Adams:2003zg,Voloshin:2008dg}.
Under the assumption that particle $c$ is correlated with particles
$\alpha$ and $\beta$ only via common correlation to the reaction
plane, one has: 
\be
\la \cos(\phi_a +\phi_\beta -2\phi_c) \ra 
=
\la \cos(\phi_a +\phi_\beta -2\psirp) \ra \, v_{2,c}, 
\label{e3p}
\ee  
where $v_{2,c}$ is the elliptic flow value of the particle $c$.
We check this assumption by using particles $c$ from different
detectors that exhibit different elliptic flow.
Figure~\ref{fig:au200F} compares the three-particle correlator
divided  by  $v_{2,c}$,
when the  third particle is
selected  from the  main  TPC with those when it is selected
from  the Forward TPCs, where elliptic flow is significantly 
smaller~\cite{Adams:2004bi}.
The shaded  bands in Fig.~\ref{fig:au200F}
and below  
illustrate the systematic error due to uncertainty in $v_2$ measurements.  
The upper (in magnitude) limit is obtained with $v_2\{4\}$
 and the lower limit from $v_2\{2\}$, the mid-point is calculated using
$v_2\{FTPC\}$~\cite{Adams:2004bi,Voloshin:2007af}.             
Whenever $v_2\{4\}$ values are  not available 
the upper limits are obtained 
assuming that the measurements with FTPC suppress 
only 50\% of the non-flow contribution.  

A very good agreement between the same
charge correlations in Fig.~\ref{fig:au200F}
justifies the assumption Eq.~\ref{e3p}.
But the opposite charge correlations are too small in magnitude 
to conclude on validity of the factorization.
Similarly, in the most peripheral collisions, the statistical
errors are large, which also prohibits making a definite conclusion.
In these cases, the factorization can be broken due to contribution 
of three-particle clusters, which we estimate with the help of
event generators.
We proceed below assuming 
 $\corr   = \mean{\cos(\phia +\phib  -2\phi_c)}/v_{2,c} $
but indicate in all plots the HIJING~\cite{refHIJING} 
direct three-particle correlation results. 
The latter can be considered as an estimate of the systematic uncertainty
from correlations not related to the reaction plane. 
(Using the event generator UrQMD yields 
both opposite-charge and same-charge correlations at least a factor
of two lower than those predicted by HIJING.)
Note that in principle such uncertainty can be
suppressed by taking particle $c$ farther in rapidity from the 
 particles $\alpha$ and $\beta$.

\begin{figure}[p]
\begin{minipage}[t]{0.5\textwidth}
 \includegraphics[width=1.1\textwidth]{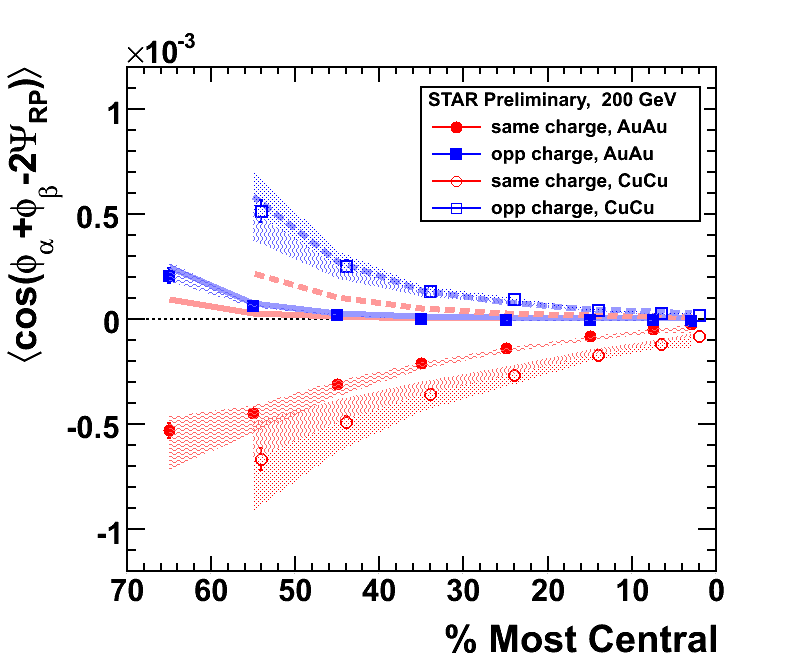}
\centerline{(a)}
\end{minipage}
\hspace{0.0\textwidth}
\begin{minipage}[t]{0.5\textwidth}
 \includegraphics[width=1.1\textwidth]{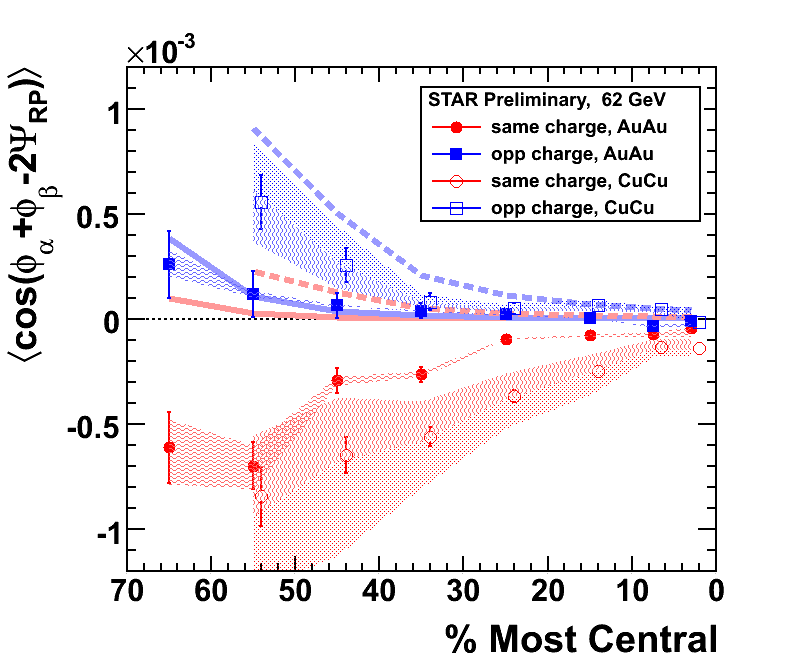}
\centerline{(b)}
\end{minipage}
 \caption{ $\mean{\cos(\phi_a +\phi_\beta -2\psirp) }$
in Au+Au and Cu+Cu
collisions at (a) $\sqrt{s_{NN}}=200$~GeV and (b)  $\sqrt{s_{NN}}=62$~GeV. 
 Thick solid (Au+Au) and dashed (Cu+Cu) 
 lines represent possible non-reaction-plane
  dependent contribution from many-particle clusters as estimated by HIJING. 
}
 \label{fig:uuv2}
\end{figure}

%==================================================
{\bf \textsl{4. Results.}}
Figure~\ref{fig:uuv2}a presents the correlator $\corr$ for Au+Au and Cu+Cu
collisions at \snn=200~GeV.
$(+,+)$  and $(-,-)$ correlations are found
to be the same within statistical errors
and are combined together as same-charge correlations.
Opposite-charge correlations are relatively smaller than same-charge
correlations. This observation led to the proposal that back-to-back
correlations may be suppressed due to the opacity of the medium as
discussed in the introduction.
The difference in magnitude between same and opposite sign correlations
is considerably smaller in Cu+Cu than in Au+Au,
qualitatively in agreement with the scenario of stronger suppression 
of the back-to-back correlations in Au+Au collisions.
In Fig.~\ref{fig:uuv2}a and below, error bars 
show the statistical errors.
Note again that presenting the results in this section we assume the
factorization of correlator, Eq.~\ref{e3p}.  
The possible error due to this assumption 
is denoted by the thick lines in Fig.~\ref{fig:uuv2}
and subsequent figures.
Figure~\ref{fig:uuv2}b shows results for collisions at 
\snn=62.4~GeV.
The signal is similar in magnitude, with slightly more pronounced 
opposite-charge correlations compared to those at \snn=200~GeV. 
This is consistent with weaker suppression of opposite-charge correlations 
in the less dense 62~GeV system.

\begin{figure}[p]
\begin{minipage}[t]{0.55\textwidth}
 \includegraphics[width=1.02\textwidth]{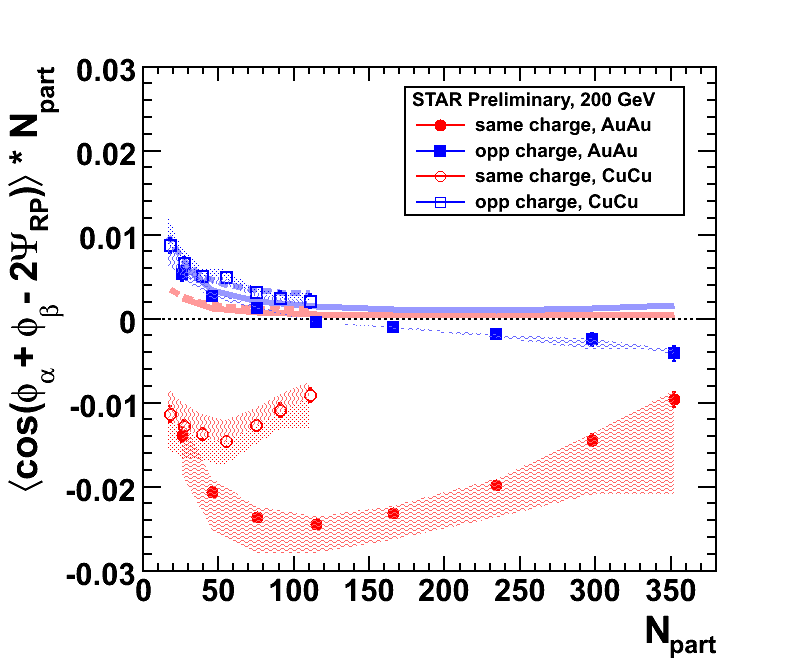}
\end{minipage}
\hspace{0.0\textwidth}
\begin{minipage}[b]{0.45\textwidth}
 \caption{
 Au+Au and Cu+Cu collisions at \snn=200~GeV. The correlations are  
scaled with the number of participants and shown as a
function number of participants.
Thick lines as in Fig.~\ref{fig:uuv2} 
\vspace*{3mm}
}
 \label{fig:uuv2_AuNpart}
\end{minipage}
\end{figure}

The correlations are weaker in more central collisions compared to more
peripheral collisions, which partially can 
be attributed to dilution of correlations
which occurs in the case of particle production from multiple sources. 
The somewhat stronger correlations in Cu+Cu collisions
than in Au+Au for the same centrality also may have similar
explanation.
To compensate for the dilution effect we show in Fig.~\ref{fig:uuv2_AuNpart}  
results multiplied by the number of participants. 
The decrease of the correlations in most central collisions 
is expected as the magnetic field weakens. 
The same and opposite sign correlations
clearly exhibit very different behavior. 
The opposite sign correlations in Au+Au and Cu+Cu collisions are
found to be very close at similar values of $N_{part}$ in rough
qualitative agreement with the picture in which their values are mostly
determined by the suppression of back-to-back correlations.

\begin{figure}[tbp]
\begin{minipage}[t]{0.5\textwidth}
 \includegraphics[width=1.08\textwidth]{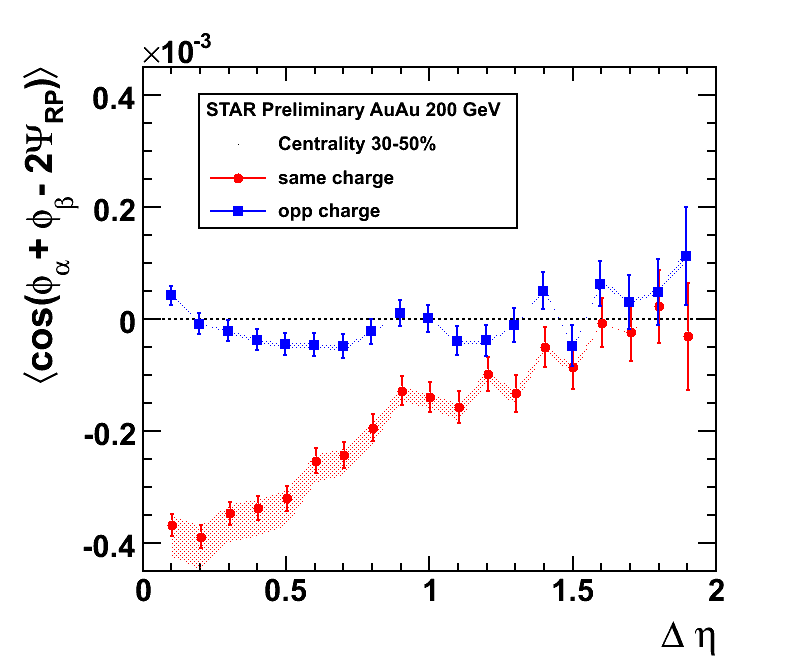} 
	\centerline{(a) }
\end{minipage}
\hspace{0.0\textwidth}
\begin{minipage}[t]{0.5\textwidth}
 \includegraphics[width=1.08\textwidth]{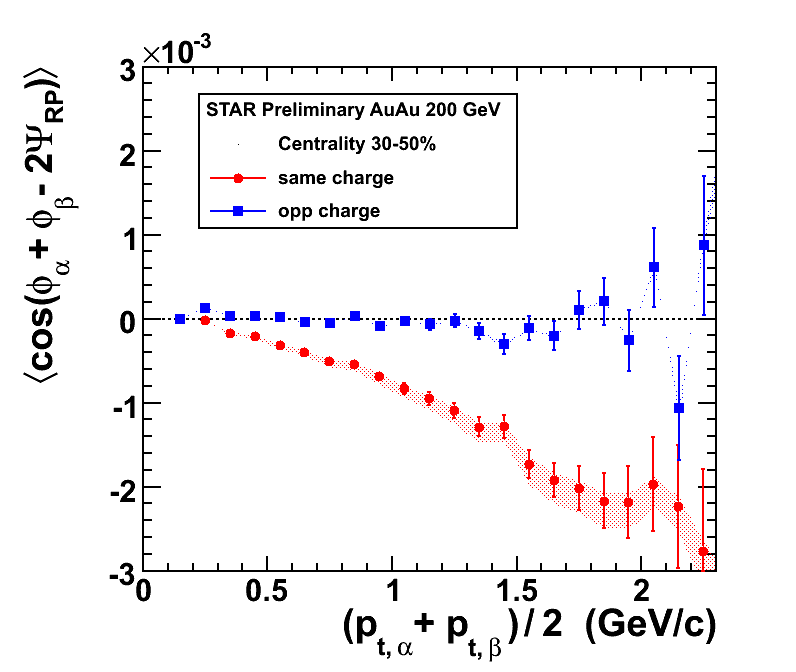} 
	\centerline{(b) }
\end{minipage}
 \caption{Au+Au at 200 GeV. The correlations dependence 
(a) on pseudorapidity separation
$\Delta\eta=|\eta_\alpha -\eta_\beta|$, and 
(b) on sum of the magnitude of transverse momenta.
} 
\label{fig:uuv2diff}
\end{figure}

Figure~\ref{fig:uuv2diff}a shows the dependence of the signal on the
difference in pseudorapidities of two particles,
 $\Delta \eta = |\eta_\alpha -\eta_\beta|$, for 30-50\% centrality.
The signal has a typical hadronic width of about one unit of
pseudorapidity.
Figure~\ref{fig:uuv2diff}b shows the dependence of the signal on 
the sum of the transverse momentum (magnitudes)  of the two particles for
these same centralities.
The signal is not concentrated in the low $p_t$
region as naively might be expected for \P-violation effects. 
We find also that the correlation depends very weakly on
$|p_{t,\alpha}-p_{t,\beta}|$ (not shown). 
This  excludes quantum interference  or
Coulomb effects as possible explanations for the signal. 
There are no specific theoretical predictions on this dependence, 
though naively one would expect that the
signal should not extend to large values of
$|p_{t,\alpha}-p_{t,\beta}|$.

%------------------------------------------------

{\bf \textsl{5. Physics background.}}
The correlator $\mean{\cos(\phia+\phib-2\psirp)}$ is a \P-even 
observable and can exhibit a non-zero signal for effects not 
related to \P-violation.
Among those are processes in which particles $\alpha$ and
$\beta$ are products of a cluster (e.g. resonance, jet,
 di-jets) decay, and the cluster itself exhibits elliptic
flow~\cite{Adams:2003zg} or decays (fragments)
differently when emitted in-plane compared to out-of-plane.
If ``flowing clusters'' are the only contribution to the
 correlator, one can write:
\be  
\hspace{-1cm}
\la \cos(\phi_\alpha + \phi_\beta -2\psirp) \ra =
\frac{ N_{\frac{clust}{event}} N_{\frac{pairs}{clust}}} 
{ N_{\frac{pairs}{event}}}    
          \,
\la \cos(\phi_\alpha + \phi_\beta -2\phi_{clust}) \ra_{clust}
\; v_{2,clust},
\label{eq:resonance}
\ee 
where $\la ... \ra_{clust}$ indicates that the average is performed only
over pairs consisting of two daughters from the same cluster.
Estimates of the contribution of ``flowing resonances'',
based on Eq.~\ref{eq:resonance} and reasonable values of resonance
abundances and values of elliptic flow, indicate that they should not
produce a  fake signal.

\begin{figure}[tbp]
\begin{minipage}[t]{0.55\textwidth}
\centerline{ \includegraphics[width=1.04\textwidth]{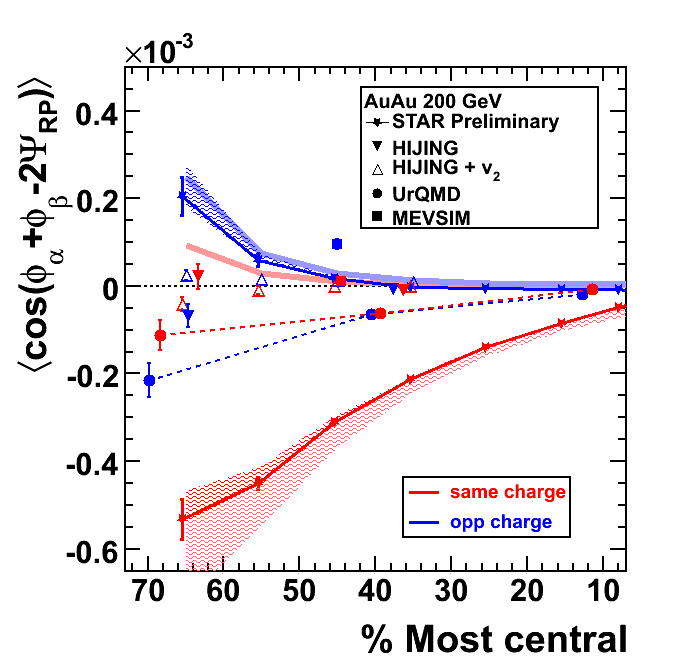}}
\end{minipage}
\begin{minipage}[b]{0.45\textwidth}
 \caption{Data comparison to simulation results for 200~GeV Au+Au.
  Blue symbols mark  {opposite-charge}
  correlations, and red are  {same-charge}.  
  Markers connected by solid lines
  represent the data.
  Acceptance cuts of $0.15<p_t<2$~GeV/c and $|\eta|<1.0$ were used
  in all cases. In simulations the true 
  reaction plane from the generated event was used.
  Thick solid lighter colored lines represent non reaction-plane
 dependent contribution as estimated by HIJING. 
 Corresponding estimates from UrQMD are about factor of two smaller.
\vspace*{3mm}
}
 \label{fig:AuAusimulations}
\end{minipage}
\end{figure}

To study the background contribution
in greater detail we used event generators MEVSIM~\cite{refMEVSIM},
 UrQMD~\cite{refRQMD} and HIJING~\cite{refHIJING}.
The results are presented in Fig.~\ref{fig:AuAusimulations}. 
MEVSIM includes only correlations due to
elliptic flow and 
resonance ($\phi$, $\Delta$, $\rho$,
$\omega$, and $K^{*}$) decay.
HIJING is also run with an added
``afterburner'' to add elliptic flow as experimentally observed.
No generator gives qualitative agreement with the data; the
model 
values of $\corr$ are significantly smaller in magnitude than what is
seen in the data, and the correlations calculated in these
models tend  to be very  similar for same and  opposite sign
correlations. 
In Fig.~\ref{fig:AuAusimulations} UrQMD points are connected by dashed
lines to illustrate that the ``reference line'' for strong parity
correlations might be not at zero. 

Here, we also mention that such effects as directed flow fluctuations and 
global polarization~\cite{Liang:2004ph,Voloshin:2004ha} should not 
have any significant contribution to the measurements.
 
%=================================================
{\bf \textsl{6. Future directions.}}  
Taking into account the importance of the question,
one can envision a dedicated program for establishing the
nature of the signal and further detail study. 
From the theoretical point of view, the calculation of the dependence
on centrality
and system system size looks fully doable though requires
significant computing and man power (e.g. 3d hydrodynamics is needed for
the calculation of the magnetic field).
Detailed predictions on the transverse momentum and particle type
dependence of the effect also will be essential in differentiating 
it from possible ``background'' contributions.
A good  theoretical understanding of the ``background'' correlations
itself is also required, as at present all event generators lack a 
good description of correlation results.

A number of future experiments and analyses are naturally suggested by
STAR results. 
One of them is the dependence of the signal on the energy of 
the colliding ions, which can be addressed, for example,
during the RHIC beam energy scan
The charge separation effect is expected to depend strongly on the formation 
of a quark-gluon plasma~\cite{Kharzeev:2007jp}, 
and the signal might be greatly suppressed or completely absent
at an energy below that at which a quark-gluon plasma can be formed.
Identified and multiparticle correlations studies also
will be available with larger statistics.
Those will be important to test a specific predictions, e.g. such as
topological cluster decays in equal number of $q\bar{q}$-pairs of all
flavors.     
The charge separation dependence on the magnetic
field~\cite{Kharzeev:2007jp} can be tested with collision of isobaric
nuclei, such as  $^{96}_{44}Ru$ and    $^{96}_{40}Zr$ that were 
used at GSI~\cite{Hong:2001tm}. Collision of isobaric nuclei will be
also very interesting in relation to the directed flow studies 
(in the latter case one needs asymmetric collisions).   

Correlation measurements from RHIC~\cite{Alver:2007wy}
and earlier measurements at ISR (see review~\cite{Foa:1975eu})
indicate that cluster formation plays an important role in multiparticle
production at high energies.
These clusters, with a size inferred in ~\cite{Alver:2007wy} to
be 2.5--3 charged particles per cluster,
may account for production of a significant fraction of all particles.
It will be interesting and important to establish how these clusters 
are related to the topological clusters as suggested 
in~\cite{Kharzeev:2001ev} and~\cite{Ostrovsky:2002cg} (``turning points''). 
Note recent progress in describing of the soft  Pomeron as a 
multi-instanton ladder~\cite{Kharzeev:2000ef}, which also suggests an
important role played in multiparticle production by topologically
nontrivial gluonic configurations.

%===========================================================
{\bf \textsl{7. Summary.}}
Charge separation due to quark interaction with topologically
 non-trivial gluonic configurations in a strong magnetic field
of non-central heavy ion collision may provide a unique opportunity
 for a direct observation of the topological structure of QCD.
The predictions are well within reach of the experiment. 
STAR has performed analysis 
of Au+Au and Cu+Cu collisions at $\sqrt{s_{NN}}$=200
and 62~GeV using three-particle correlations that are directly
sensitive to the local \P-violation effects in heavy-ion collisions.
The results are reported for different particle charge combinations 
as a function of collision centrality, particle
separation in pseudorapidity, and particle transverse momentum.
Qualitatively the results agree 
with the magnitude and gross features
of the theoretical predictions for local \P-violation in heavy-ion
collisions, but the signal persists to higher transverse
momenta than expected~\cite{Kharzeev:2007jp}.
The particular observable used in the analysis is
\P-even and might be sensitive to non-parity-violating effects. 
None of
the studies we have performed so far have revealed a background source
that can explain the observed same-sign correlations.

Better theoretical calculations of the
expected signal and potential physics backgrounds 
in high energy heavy-ion collisions  are essential for confirmation  and 
experimental study of this phenomenon. 

A very exciting future program dedicated to detail study of the effect
is emerging. 

{\bf \textsl{Acknowledgments.}}  
Discussions with D.~Kharzeev and E.~Shuryak are gratefully 
acknowledged.

%-------------------------------------------------------
\medskip
%\noindent

\end{document}